\documentclass[11pt]{article}
\usepackage[dvips]{graphicx}
\usepackage{float}
\usepackage{epsfig}
\usepackage{latexsym,amsmath,amsfonts,amssymb}
\usepackage{rotating}
\usepackage[american]{babel}
\usepackage[dvips]{graphicx}
\usepackage{bbm}
\usepackage{color}
\usepackage{slashed}
\usepackage[unicode]{hyperref}
\usepackage{lscape}
\usepackage{bigints}
\usepackage{enumerate}
\usepackage[shortlabels]{enumitem}
\usepackage{subcaption}
\usepackage{tikz}
\usetikzlibrary{decorations.pathreplacing}
\usetikzlibrary{shapes}
\usepackage{graphicx}
\usepackage{epsfig}
\usepackage{rotating}
\usepackage{amssymb}
\usepackage{dsfont}
\usepackage{psfrag}
\usepackage{amsmath,euscript,array,mathrsfs,amsfonts}
\usepackage{slashed}
\usepackage{array}
\usepackage{youngtab}
\usepackage{color}
\usepackage{bbold}
\usepackage{hyperref}
\usepackage[thinc]{esdiff}
\hypersetup{
colorlinks = true,
linkcolor = red,
linktocpage = true,
citecolor = blue
}

\usepackage{tikz}
\usetikzlibrary{calc}
 \usetikzlibrary{decorations.text}
 \usetikzlibrary{shapes}

 \usetikzlibrary{decorations.pathmorphing}
\usetikzlibrary{decorations.pathreplacing}
\usetikzlibrary{arrows.meta}
\tikzset{
  >={To[length=5pt]}
  }
\usetikzlibrary{shapes, shapes.geometric, shapes.symbols, shapes.arrows, shapes.multipart, shapes.callouts, shapes.misc}
\tikzset{snake it/.style={decorate, decoration=snake}}
\tikzset{7brane/.style={circle, draw=black, fill=black,ultra thick,inner sep=1.5 pt, minimum size=1 pt,}, c/.default={4pt}}
\tikzset{cross/.style={cross out, draw=black,thick, minimum size=2*(#1-\pgflinewidth), inner sep=0pt, outer sep=0pt}, cross/.default={5pt}}
\tikzset{big7brane/.style={circle, draw=black, fill=black,ultra thick,inner sep=2.5 pt, minimum size=1 pt,}, c/.default={4pt}}
\tikzset{u/.style={circle, draw=black, fill=white,inner sep=2 pt, minimum size=2 pt,},f/.style={square, draw=black, fill=white,ultra thick,inner sep=4 pt, minimum size=2 pt,}}
\tikzset{so/.style={circle, draw=black, fill=red,inner sep=2 pt, minimum size=2 pt,},f/.style={square, draw=black, fill=white,ultra thick,inner sep=4 pt, minimum size=2 pt,}}
\tikzset{sp/.style={circle, draw=black, fill=blue,inner sep=2 pt, minimum size=2 pt,},f/.style={square, draw=black, fill=white,ultra thick,inner sep=4 pt, minimum size=2 pt,}}
\tikzset{uf/.style={rectangle, draw=black, fill=white,inner sep=3 pt, minimum size=4 pt,}}
\tikzset{spf/.style={rectangle, draw=black, fill=blue, thick,inner sep=3 pt, minimum size=4 pt, circle, draw=black, fill=blue,thick,inner sep=2 pt, minimum size=2 pt,},f/.style={square, draw=black, fill=white,ultra thick,inner sep=4 pt, minimum size=2 pt,}}
\tikzset{sof/.style={rectangle, draw=black, fill=red, thick,inner sep=3 pt, minimum size=4 pt,}}
\usetikzlibrary{positioning}
\usetikzlibrary{arrows}
\usetikzlibrary{decorations.pathreplacing}
\usetikzlibrary{shapes}

\makeatletter\def\l@subsubsection#1#2{}%
\makeatother

\renewcommand\theequation{\arabic{section}.\arabic{equation}} 

\def\cA{{\cal A}}

\def\cF{{\cal F}}

\def\CC{\ensuremath{\mathds C}}
\def\RR{\ensuremath{\mathds R}}
\def\ZZ{\ensuremath{\mathds Z}}

\DeclareMathOperator{\vol}{vol}

\DeclareMathOperator{\sech}{sech}
\DeclareMathOperator{\tr}{tr}

\DeclareMathOperator{\csch}{csch}

\newcommand{\be}{\begin{equation}}
\newcommand{\ee}{\end{equation}}
\newcommand{\ba}{\begin{eqnarray}}
\newcommand{\ea}{\end{eqnarray}}

\def\im{Invent. Math.}

\def\hat{\widehat}
\def\a{\alpha}
\def\b{\beta}
\def\c{\gamma}
\def\d{\delta}
\def\f{\phi}               
\def\vf{\varphi}  
\def\tvf{\tilde{\varphi}}
\def\vp{\varphi}
\def\g{\gamma}
\def\h{\eta}
\def\j{\psi}
\def\k{\kappa}                    
\def\l{\lambda}
\def\m{\mu}
\def\n{\nu}
\def\o{\omega}  \def\w{\omega}
\def\q{\theta}  \def\th{\theta}                  
\def\r{\rho}                                     
\def\s{\sigma}                                   
\def\t{\tau}
\def\u{\upsilon}
\def\x{\xi}
\def\z{\zeta}
\def\pt{\tilde{\varphi}}
\def\tt{\tilde{\theta}}
\def\lab{\label}
\def\wg{\wedge}
\def\bpsi{\bar{\psi}}
\def\bt{\bar{\theta}}
\def\bvf{\bar{\varphi}}
\def\W{\Omega}

\newcommand{\nb}{\nonumber}
\newcommand{\td}{\mathrm{d}}

\DeclareMathOperator{\str}{str}

\newcommand{\beq}{\begin{equation}}
\newcommand{\eeq}{\end{equation}}
\newcommand{\bea}{\begin{eqnarray}}
\newcommand{\eea}{\end{eqnarray}}

\newcommand{\beqs}{\begin{eqnarray}}
\newcommand{\eeqs}{\end{eqnarray}}
\newcommand{\bal}{\begin{aligned}}
\newcommand{\eal}{\end{aligned}}
\makeatletter
\newcommand\setItemnumber[1]{\setcounter{enum\romannumeral\@enumdepth}{\numexpr#1-1\relax}}
\makeatother
\begin{document}
\baselineskip=15.5pt
\pagestyle{plain}
\setcounter{page}{1}

\def\del{{\partial}}
\def\vev#1{\left\langle #1 \right\rangle}
\def\cn{{\cal N}}
\def\co{{\cal O}}


\def\IC{{\mathbb C}}
\def\IR{{\mathbb R}}
\def\IZ{{\mathbb Z}}
\def\RP{{\bf RP}}
\def\CP{{\bf CP}}
\def\Poincaré{{Poincar\'e }}
\def\tr{{\rm tr}}
\def\tp{{\tilde \Phi}}

\def\TL{\hfil$\displaystyle{##}$}
\def\TR{$\displaystyle{{}##}$\hfil}
\def\TC{\hfil$\displaystyle{##}$\hfil}
\def\TT{\hbox{##}}
\def\HLINE{\noalign{\vskip1\jot}\hline\noalign{\vskip1\jot}}
\def\seqalign#1#2{\vcenter{\openup1\jot
   \halign{\strut #1\cr #2 \cr}}}
\def\lbldef#1#2{\expandafter\gdef\csname #1\endcsname {#2}}
\def\eqn#1#2{\lbldef{#1}{(\ref{#1})}%
\begin{equation} #2 \label{#1} \end{equation}}
\def\eqalign#1{\vcenter{\openup1\jot
     \halign{\strut\span\TL & \span\TR\cr #1 \cr
    }}}

\def\eno#1{(\ref{#1})}
\def\href#1#2{#2}
\def\half{\frac{1}{2}}



\def\ads{{\it AdS}}
\def\adsp{{\it AdS}$_{p+2}$}
\def\cft{{\it CFT}}

\newcommand{\ber}{\begin{eqnarray}}
\newcommand{\eer}{\end{eqnarray}}

\newcommand{\beqar}{\begin{eqnarray}}
\newcommand{\cO}{{\cal O}}
\newcommand{\cT}{{\cal T}}
\newcommand{\cR}{{\cal R}}
\newcommand{\eeqar}{\end{eqnarray}}
\newcommand{\tht}{\thteta}
\newcommand{\lm}{\lambda}\newcommand{\Lm}{\Lambda}


\newcommand{\nonu}{\nonumber}
\newcommand{\oh}{\displaystyle{\frac{1}{2}}}
\newcommand{\dsl}
   {\kern.06em\hbox{\raise.15ex\hbox{$/$}\kern-.56em\hbox{$\partial$}}}
\newcommand{\as}{\not\!\! A}
\newcommand{\ps}{\not\! p}
\newcommand{\ks}{\not\! k}
\newcommand{\D}{{\cal{D}}}
\newcommand{\dv}{d^2x}
\newcommand{\Z}{{\cal Z}}
\newcommand{\N}{{\cal N}}
\newcommand{\Dsl}{\not\!\! D}
\newcommand{\Bsl}{\not\!\! B}
\newcommand{\Psl}{\not\!\! P}

\newcommand{\eeqarr}{\end{eqnarray}}


\def\del{{\delta^{\hbox{\sevenrm B}}}} \def\ex{{\hbox{\rm e}}}
\def\azb{A_{\bar z}} \def\az{A_z} \def\bzb{B_{\bar z}} \def\bz{B_z}
\def\czb{C_{\bar z}} \def\cz{C_z} \def\dzb{D_{\bar z}} \def\dz{D_z}
\def\im{{\hbox{\rm Im}}} \def\mod{{\hbox{\rm mod}}} \def\tr{{\hbox{\rm Tr}}}
\def\ch{{\hbox{\rm ch}}} \def\imp{{\hbox{\sevenrm Im}}}
\def\trp{{\hbox{\sevenrm Tr}}} \def\vol{{\hbox{\rm Vol}}}
\def\rl{\Lambda_{\hbox{\sevenrm R}}} \def\wl{\Lambda_{\hbox{\sevenrm W}}}
\def\fc{{\cal F}_{k+\cox}} \def\vev{vacuum expectation value}
\def\nodiv{\mid{\hbox{\hskip-7.8pt/}}}
\def\ie{{\em i.e.}}
\def\ie{\hbox{\it i.e.}}

\def\CC{{\mathchoice
{\rm C\mkern-8mu\vrule height1.45ex depth-.05ex
width.05em\mkern9mu\kern-.05em}
{\rm C\mkern-8mu\vrule height1.45ex depth-.05ex
width.05em\mkern9mu\kern-.05em}
{\rm C\mkern-8mu\vrule height1ex depth-.07ex
width.035em\mkern9mu\kern-.035em}
{\rm C\mkern-8mu\vrule height.65ex depth-.1ex
width.025em\mkern8mu\kern-.025em}}}

\def\RR{{\rm I\kern-1.6pt {\rm R}}}
\def\NN{{\rm I\!N}}
\def\ZZ{{\rm Z}\kern-3.8pt {\rm Z} \kern2pt}
\def\IB{\relax{\rm I\kern-.18em B}}
\def\ID{\relax{\rm I\kern-.18em D}}
\def\II{\relax{\rm I\kern-.18em I}}
\def\IP{\relax{\rm I\kern-.18em P}}
\newcommand{\CS}{{\scriptstyle {\rm CS}}}
\newcommand{\CSs}{{\scriptscriptstyle {\rm CS}}}
\newcommand{\rc}{\nonumber\\}
\newcommand{\bear}{\begin{eqnarray}}
\newcommand{\eear}{\end{eqnarray}}

\newcommand{\LL}{{\cal L}}

\def\mani{{\cal M}}
\def\calo{{\cal O}}
\def\calb{{\cal B}}
\def\calw{{\cal W}}
\def\calz{{\cal Z}}
\def\cald{{\cal D}}
\def\calc{{\cal C}}
\newcommand{\gt}{\tilde{g}}

\def\to{\rightarrow}
\def\ele{{\hbox{\sevenrm L}}}
\def\ere{{\hbox{\sevenrm R}}}
\def\zb{{\bar z}}
\def\wb{{\bar w}}
\def\nodiv{\mid{\hbox{\hskip-7.8pt/}}}
\def\menos{\hbox{\hskip-2.9pt}}
\def\dr{\dot R_}
\def\drr{\dot r_}
\def\ds{\dot s_}
\def\da{\dot A_}
\def\dga{\dot \gamma_}
\def\ga{\gamma_}
\def\dal{\dot\alpha_}
\def\al{\alpha_}
\def\cl{{closed}}
\def\cls{{closing}}
\def\vev{vacuum expectation value}
\def\tr{{\rm Tr}}
\def\to{\rightarrow}
\def\too{\longrightarrow}

\newcommand{\dd}{\mathrm{d}}

\def\a{\alpha}
\def\b{\beta}
\def\c{\gamma}
\def\d{\delta}
\def\e{\epsilon}           
\def\F{\Phi}
\def\f{\phi}               
\def\vf{\varphi}  \def\tvf{\tilde{\varphi}}
\def\vp{\varphi}
\def\g{\gamma}
\def\h{\eta}
\def\j{\psi}
\def\k{\kappa}                    
\def\l{\lambda}
\def\m{\mu}
\def\n{\nu}
\def\o{\omega}  \def\w{\wedge}
\def\q{\theta}  \def\th{\theta}                  
\def\r{\rho}                                     
\def\s{\sigma}                                   
\def\t{\tau}
\def\u{\upsilon}
\def\x{\xi}
\def\X{\Xi}
\def\z{\zeta}
\def\pt{\tilde{\varphi}}
\def\tt{\tilde{\theta}}
\def\lab{\label}
\def\6{\partial}
\def\wg{\wedge}
\def\atanh{{\rm arctanh}}
\def\bpsi{\bar{\psi}}
\def\bt{\bar{\theta}}
\def\bvf{\bar{\varphi}}

\def\ft#1#2{{\textstyle{{\scriptstyle #1}\over {\scriptstyle #2}}}}
\def\fft#1#2{{#1 \over #2}}
\def\del{\partial}
\def\sst#1{{\scriptscriptstyle #1}}

\def\dalemb#1#2{{\vbox{\hrule height .#2pt
        \hbox{\vrule width.#2pt height#1pt \kern#1pt
                \vrule width.#2pt}
        \hrule height.#2pt}}}
\def\square{\mathord{\dalemb{6.8}{7}\hbox{\hskip1pt}}}
\def\hF{\hat F}
\def\tA{\widetilde A}
\def\tcA{{\widetilde{\cal A}}}
\def\tcF{{\widetilde{\cal F}}}
\def\hA{\hat{\cal A}}
\def\cF{{\cal F}}
\def\cA{{\cal A}}
\def\wdg{{\sst \wedge}}

\def\0{{\sst{(0)}}}
\def\1{{\sst{(1)}}}
\def\2{{\sst{(2)}}}
\def\3{{\sst{(3)}}}
\def\4{{\sst{(4)}}}
\def\5{{\sst{(5)}}}
\def\6{{\sst{(6)}}}
\def\7{{\sst{(7)}}}
\def\8{{\sst{(8)}}}
\def\n{{\sst{(n)}}}
\def\tV{\widetilde V}
\def\tW{\widetilde W}
\def\tH{\widetilde H}
\def\tE{\widetilde E}
\def\tF{\widetilde F}
\def\tA{\widetilde A}
\def\tP{{\widetilde P}}
\def\tD{\widetilde D}
\def\bA{\bar{\cal A}}
\def\bF{\bar{\cal F}}
\def\tG{\widetilde G}
\def\tT{\widetilde T}
\def\Z{\rlap{\sf Z}\mkern3mu{\sf Z}}
\def\R{\rlap{\rm I}\mkern3mu{\rm R}}
\def\G{{\Gamma}}
\def\gg{\bf g}
\def\CS{{\cal S}}
\def\S{{\cal S}}
\def\P{{\cal P}}
\def\ep{\epsilon}
\def\td{\tilde}
\def\wtd{\widetilde}
\def\half{{\textstyle{1\over2}}}
\def\Qw{{Q_{\rm wave}}}
\def\Qnut{{Q_{\sst{\rm NUT}}}}
\def\mun{{\mu_{\sst{\rm NUT}}}}
\def\muw{{\mu_{\rm wave}}}
\let\a=\alpha \let\b=\beta \let\g=\gamma \let\d=\delta \let\e=\epsilon
\let\z=\zeta \let\h=\eta \let\q=\theta \let\i=\iota \let\k=\kappa
\let\l=\lambda \let\m=\mu \let\n=\nu \let\x=\xi 
\let\s=\sigma \let\t=\tau \let\u=\upsilon \let\f=\phi \let\c=\chi \let\y=\psi
\let\w=\omega  \let\D=\Delta \let\Q=\Theta \let\L=\Lambda
\let\X=\Xi  \let\U=\Upsilon \let\F=\Phi \let\Y=\Psi
\let\C=\Chi \let\W=\Omega     
\let\la=\label \let\ci=\cite \let\re=\ref
\let\se=\section \let\sse=\subsection \let\ssse=\subsubsection 
\def\bd{\begin{document}} \def\ed{\end{document}}
\def\ds{\documentstyle} \let\fr=\frac \let\bl=\bigl \let\br=\bigr
\let\Br=\Bigr \let\Bl=\Bigl 
\let\bm=\bibitem
\let\na=\nabla
\let\pa=\partial \let\ov=\overline 
\def\ba{\begin{eqnarray}}
\def\ea{\end{eqnarray}}
\def\ft#1#2{{\textstyle{{\scriptstyle #1}\over {\scriptstyle #2}}}}
\def\fft#1#2{{#1 \over #2}}
\def\del{\partial}
\def\sst#1{{\scriptscriptstyle #1}}
\def\oneone{\rlap 1\mkern4mu{\rm l}}
\def\ie{{\it i.e.\ }}
\def\via{{\it via}}
\def\semi{{\ltimes}}
\def\str{{\rm str}}
\def\jm{{\rm j}}
\def\im{{\rm i}}
\def\mapright#1{\smash{\mathop{-\!\!\!-\!\!\!-\!\!\!-\!\!\!-\!\!\!
             \longrightarrow}\limits^{#1}}}
\def\maprightt#1#2{\smash{\mathop{-\!\!\!-\!\!\!-\!\!\!-\!\!\!-\!\!\!
             \longrightarrow}\limits^{#1}_{#2}}}

\newcommand{\ho}[1]{$\, ^{#1}$}
\newcommand{\hoch}[1]{$\, ^{#1}$}
\newcommand{\ra}{\rightarrow}
\newcommand{\lra}{\longrightarrow}
\newcommand{\Lra}{\Leftrightarrow}
\newcommand{\bp}{\tilde \beta^\prime}
\newcommand{\Tr}{{\rm Tr} } 
\def\rme{{\rm e}}


\newfont{\namefont}{cmr10}
\newfont{\addfont}{cmti7 scaled 1440}
\newfont{\boldmathfont}{cmbx10}
\newfont{\headfontb}{cmbx10 scaled 1728}

\newcommand{\hyph}[1]{$#1$\nobreakdash-\hspace{0pt}}
\providecommand{\abs}[1]{\lvert#1\rvert}
\newcommand{\Nugual}[1]{$\mathcal{N}= #1 $}
\newcommand{\sub}[2]{#1_\text{#2}}
\newcommand{\partfrac}[2]{\frac{\partial #1}{\partial #2}}
\newcommand{\bsp}[1]{\begin{equation} \begin{split} #1 \end{split} \end{equation}}
\newcommand{\calF}{\mathcal{F}}
\newcommand{\calO}{\mathcal{O}}
\newcommand{\calM}{\mathcal{M}}
\newcommand{\calV}{\mathcal{V}}
\newcommand{\bbZ}{\mathbb{Z}}
\newcommand{\bbC}{\mathbb{C}}
\newcommand{\cK}{{\cal K}}

\newcommand{\Thq}{\Theta\left(\r-\r_q\right)}
\newcommand{\Dq}{\d\left(\r-\r_q\right)}
\newcommand{\kten}{\kappa^2_{\left(10\right)}}
\newcommand{\pbi}[1]{\imath^*\left(#1\right)}
\newcommand{\tth}{\tilde{\th}}
\newcommand{\tf}{\tilde{\f}}
\newcommand{\tj}{\tilde{\j}}
\newcommand{\tw}{\tilde{\omega}}
\newcommand{\tz}{\tilde{z}}
\newcommand{\prj}[2]{(\partial_r{#1})(\partial_{\j}{#2})-(\partial_r{#2})(\partial_{\j}{#1})}
\def\atanh{{\rm arctanh}}
\def\sech{{\rm sech}}
\def\csch{{\rm csch}}
\allowdisplaybreaks[1]

\def\red{\textcolor[rgb]{0.98,0.00,0.00}}

\newcommand{\Dan}[1] {{\textcolor{blue}{#1}}}

\numberwithin{equation}{section}



%

%
\setcounter{footnote}{0}
\renewcommand{\theequation}{{\rm\thesection.\arabic{equation}}}

\begin{titlepage}

\begin{center}

\vskip .5in 
\noindent

{\Large \bf{ 
TsT-Generated Solutions in Type IIB Supergravity from Twisted Compactification of AdS$_5\times$T$^{1,1}$
}}
\bigskip\medskip

Federico Castellani$^\dagger$\footnote{federico.castellani@unifi.it}

\bigskip\medskip
{\small 
$^\dagger$INFN, Sezione di Firenze and
Dipartimento di Fisica e Astronomia, Universit\'a di Firenze, 

Via G. Sansone 1, I-50019 Sesto Fiorentino (Firenze), Italy.
}

\vskip .5cm 
\vskip .9cm 
     	{\bf Abstract }\vskip .1in
\end{center}

\noindent
This paper investigates marginal and dipole TsT transformations of a seed type IIB supergravity solution dual to a supersymmetry-preserving deformation of the Klebanov-Witten 4d SCFT. To explore key properties of the deformed theories, we holographically analyze various observables, including Wilson loops, 't Hooft loops, Entanglement Entropy, and holographic central charge flow. Moreover, we focus on detecting which of these observables are affected by the dynamics of the Kaluza-Klein (KK) modes resulting from the circle compactification.
 \noindent
\vskip .5cm
\vskip .5cm
\vfill
\eject

\end{titlepage}

\setcounter{footnote}{0}

\small{
\tableofcontents}

\normalsize

\newpage
\renewcommand{\theequation}{{\rm\thesection.\arabic{equation}}}
%
\section{Introduction and general idea}
The AdS/CFT conjecture and its refinements \cite{Maldacena:1997re, Gubser:1998bc, Witten:1998qj}  provide a useful tool allowing the study of strongly coupled systems through their gravitational duals.
Various extensions of the original formulation of the correspondence have explored the idea of applying the gauge/gravity duality in non-conformal systems, see \textit{e.g.} \cite{Itzhaki:1998dd,Witten:1998zw,Boonstra:1998mp}. 
Along these lines, one possible extension leading to the study of non-conformal and less supersymmetric (SUSY) quantum field theories goes through wrapped brane constructions \textit{e.g.} \cite{Maldacena:2000yy,Atiyah:2000zz, Girardello:1999hj,Polchinski:2000uf}. Moreover, an alternative approach is based on D$_p$-brane systems on conifolds returning a geometrical description for non-perturbative features of quiver theories, see \textit{e.g.} \cite{Klebanov:1998hh, Klebanov:2000nc,Klebanov:2000hb,Gubser:2004qj}.\\
Recently, novel analysis focused on supersymmetric twisted circle compactification in backgrounds belonging to these two families have been presented \cite{Nunez:2023nnl,Nunez:2023xgl,Fatemiabhari:2024aua,Kumar:2024pcz,Chatzis:2024kdu,Chatzis:2024top}. In particular, in the quantum field theories considered, circle compactification with anti-periodic boundary conditions for fermions are performed. The Bose-Fermi mass degeneracy for the Kaluza-Klein modes and some amount of SUSY are preserved via the introduction of a background gauge field for the R-symmetry to compensate for the fermions' anti-periodicity. Remarkably, the presence of a partially-preserved supersymmetry ensures the stability of the background. In the special case of 4d $\mathcal{N}=4$ Super Yang Mills (SYM) on $\mathbb{R}^{1,2}\times S^1$ a full detailed description of the four-supercharges SUSY preserving twisting deformation is provided for both the field theory and supergravity dual sides \cite{Kumar:2024pcz,Cassani:2021fyv, Chatzis:2024kdu,Chatzis:2024top}. The cases of more general Lagrangian and non-Lagrangian theories are considered in \cite{Nunez:2023nnl,Nunez:2023xgl,Fatemiabhari:2024aua,Chatzis:2024kdu,Chatzis:2024top}.\\
The compactification process introduces an infinite tower of massive Kaluza-Klein (KK) modes, displaying masses proportional to the inverse of the reduction circle radius. In the limit in which the supergravity approximation of the dual closed string theory models is reliable, however, it is not possible to decouple these KK modes from the far IR dynamics, as their masses are of the same order of the energy scale of interest. In evaluating holographically quantum field theory observables it is thus difficult to separate contributions from the strict IR sector (massless modes) and the UV completion of the system (KK massive modes).  \\
To analyze the effects of the KK dynamics on the QFT observables and distinguish them from the purely gauge contributions, a serviceable and useful approach has been presented in \cite{Gursoy:2005cn, Castellani:2024ial}.
The main idea behind this approach goes along the lines of comparing field theory observables calculated from the original (seed) holographic dual background with the same observables deduced from a carefully deformed background.
More specifically, the latter is generated via a so-called TsT transformation, which is the result of a T-duality on one circle, followed by a shift/change of coordinates and another T-duality. This class of deformations, introduced by Lunin and Maldacena in \cite{Lunin:2005jy}, modifies the original background by exploiting symmetries of the internal manifold. This transformation returns new supergravity solutions corresponding to marginal deformations of the dual field theory as, for instance, beta deformations of the superpotential in the 4d $\mathcal{N}=4$ SYM case \cite{Leigh:1995ep}. Remarkably, a TsT deformation involving a $U(1) \times U(1)$ toroidal isometry, transverse to the field theory directions, modifies only the operators charged under these symmetries. Thus, in the deformed dual field theory, we expect the dynamics of the KK modes (which are typically charged under internal symmetries) to be affected by the transformation, while the gauge theory sector remains untouched.
Therefore, if we detect differences (and dependence on the transformation parameters) in an observable when comparing its values in the seed and deformed backgrounds, we can infer that this observable receives non-trivial contributions from the KK sector.

In this paper, we extend the study of deformations of SUSY twisted circle compactified 4d $\mathcal{N}=4$ SYM in \cite{Castellani:2024ial} to investigate two specific types of deformations, \textit{i.e.} marginal and dipole ones, in the context of the Klebanov-Witten $\mathcal{N}=1$ 4d SCFT and its holographic dual \cite{Klebanov:1998hh}.  The main focuses of this work are the following.

In first place, we aim to generalize the results of \cite{Castellani:2024ial} to other type IIB supergravity solutions within the family of SUSY-preserving circle reductions, presented in \textit{e.g.}, \cite{Chatzis:2024kdu,Chatzis:2024top}. These backgrounds present challenges, particularly in disentangling IR and UV effects when evaluating field theory observables in the supergravity limit. Therefore, we focus on identifying universal features related to the dynamics of the KK modes and their impact on key system properties, such as confinement, magnetic monopole screening, and symmetry-breaking patterns.

In second place, while in the case of SUSY twisted circle compactified 4d $\mathcal{N}=4$ SYM, the reduction process is well understood both in the field theory and the dual supergravity solution, allowing a comparison between the two sides, this is not the case for the Klebanov-Witten model and $AdS_5\times T^{1,1}$. Hence, the holographic findings achieved in this paper lead to non-trivial predictions on relevant observable properties and their IR and UV sectors dependence in the context of partial supersymmetric twisted circle compactification of Klebanov-Witten $\mathcal{N}=1$ 4d SCFT.\\

The paper is organized as follows. \\
In Section \ref{section-geometry}, we provide a brief discussion of the Anabalo\'n-Ross-like deformed type II supergravity solutions. We further present a short overview of the $\widehat{AdS_5}\times \widehat{T^{1,1}}$ soliton background and its main features.
We review the TsT-generating solution procedure and discuss applications to the $\widehat{AdS_5}\times \widehat{T^{1,1}}$ solution. In particular, we study both the cases of marginal and dipole TsT transformations. In the former, the $U(1)\times U(1)$ symmetry involved in the TsT process is taken inside the isometry group of $T^{1,1}$ (R-symmetries), while in the latter if one $U(1)$ is again laying inside $T^{1,1}$, the other is taken along the QFT compactification circle. The two backgrounds are presented with details, quantized charges are discussed and (when it is possible) geometric invariants written.\\
In Section \ref{QFT-section}, we proceed to calculate within the holographic approach the observables related to the QFTs corresponding to the TsT marginal and dipole deformed background introduced in Section \ref{section-geometry}. We further comment on the similarities and differences with the results for the same quantities holographically computed from the seed solution. Among the possible observables, we examine Wilson loops and 't Hooft loops, highlighting footprints of confinement in the present systems. Furthermore, we analyze Polyakov-like loops, showing that the breaking of a $\mathbb{Z}_N$ symmetry is present. The Entanglement Entropy is calculated and discussed in both solutions.\\
We also compute the central charge flow, which reflects the number of degrees of freedom in the QFTs. This is an energy-dependent quantity and goes from zero degrees of freedom at low energy, indicating a gapped system, to the degrees of freedom of the UV CFT in the high-energy limit. In the dipole case, notably, the number of degrees of freedom diverges in the UV, suggesting a non-local, non-field theoretical description. \\
In Section \ref{Conclusions}, conclusions, closing remarks, and possible future developments are given.

\section{Geometry}\label{section-geometry}
In this section, we aim to briefly review what has been recently done in \cite{Castellani:2024ial} regarding the study of TsT-generated solutions in type IIB supergravity and to extend the analysis to alternative Anabalon-Ross-like deformations of
AdS geometries with different internal manifolds. In particular, here we focus on the type IIB background of the form 
\begin{eqnarray}
   \widehat{AdS_5}\times \widehat{T^{1,1}}\,,
\end{eqnarray}
where the hats indicate geometry deformations of $AdS_5$ and the internal manifold $T^{1,1}$.\\
Let us begin by providing the relevant properties of the family of smooth type II backgrounds shown in \textit{e.g.} \cite{Anabalon:2021tua, Chatzis:2024top,Chatzis:2024kdu, Kumar:2024pcz}. The seed-background for this set of supergravity solutions is $AdS_5\times M^{5}$, with $M^5$ being a five-dimensional smooth manifold as \textit{e.g.} $S^5$ or $T^{1,1}$.
A compactification on a $S^1$ circle of radius $R$ of one of the Poincaré directions of $AdS_5$ (that we name as the $\phi$ direction) is performed. This is followed by a deformation given by introducing a cigar-like geometry in the subspace of $AdS_5$ spanned by $\phi$ and the holographic radial direction $r$. 
This is realized through a warping function $f(r)$ at whose zero the $S^1_\phi$ circle is smoothly shirking to zero size. Moreover, to preserve at least four supercharges among the sixteen of the original supergravity solution, a fine-tuned fibration for the internal manifold $M^5$ over the $S^1_\phi$ circle direction is considered. For a detailed analysis of a set of backgrounds belonging to the family presented above, we mainly refer to \cite{Castellani:2024ial, Anabalon:2021tua, Chatzis:2024top,Chatzis:2024kdu}.\\
In the present work, as we have anticipated before, we mainly focus on the case of $\widehat{AdS_5}\times \widehat{T^{1,1}}$.
We start by displaying the main features of the five-dimensional $T^{1,1}$ manifold. The latter can be seen as an $S^1$ bundle over $S^2\times S^2$, and its line element can be expressed as
\begin{equation}
\label{T11metric}
\mathrm{d}s^2_{T^{1,1}}=\frac{1}{6}\sum_{i=1}^2\left(\mathrm{d}\theta_i^2+\sin^2\theta_i\mathrm{d}\phi_i^2\right) + \frac{1}{9}\left(\mathrm{d}\psi +\sum_{i=1}^2 \cos\theta_i\mathrm{d}\phi_i\right)^2,
\end{equation}
where we have introduced a parametrization given in terms of the four angles $(\theta_1,\phi_1,\theta_2,\phi_2)$ of the two $S^2$, and the $S^1$, $\psi$ direction. Notice that in this case $\psi$ ranges in the interval  $[0,4\pi]$. \\
Thus, following the deformation algorithm sketched above, starting from $AdS_5\times T^{1,1}$ it is possible to build up the soliton $\widehat{AdS_5}\times\widehat{T^{1,1}}$ background.
The latter deformed type IIB solution is then given by  \cite{Chatzis:2024kdu,Chatzis:2024top}
\bea
\label{AdS5xT11}
\mathrm{d}s^2_{10}&=&r^2 (-\mathrm{d}t^2+\mathrm{d}x_1^2 + \mathrm{d}x_2^2 +  f(r)\mathrm{d}\phi^2) + \frac{\mathrm{d}r^2}{ r^2 f(r)}+\nb\\
&&+\left[\frac{1}{6}\sum_{i=1}^2\left(\mathrm{d}\theta_i^2+\sin^2\theta_i\mathrm{d}\phi_i^2\right) + \frac{1}{9}\left(\mathrm{d}\psi + \sum_{i=1}^2\cos\theta_i\mathrm{d}\phi_i+3\mathcal{A}\right)^2\right]\,,
\eea
with the warping factor and fibration gauge field \footnote{In general, the warping function $f(r)$ can be defined as $f(r) = 1-\frac{\m^2}{r^4}-\frac{Q^6}{r^6}$. Only the solution with $\m=0$ is SUSY preserving. We focus mostly on the $\mu=0$ case in what follows. We also set $g_s=\alpha'=1$.}
\bea
f(r) &=& 1-\frac{Q^6}{r^6}\,,\quad \mathcal{A}= Q^3 \left( \frac{1}{r^2}- \frac{1}{Q^2}\right)\mathrm{d}\phi\,,
\eea
and $r = Q$ being the minimal radius value.
The solution is completed by the following Ramond-Ramond fluxes
\bea
\label{AdS5xT11_G5}
&&F_5=\left(1+\star_{10}\right) \,G_5\,, \qquad G_5=-4r^3\mathrm{d}t\wg \mathrm{d}x_1 \wg\mathrm{d}x_2\wg \mathrm{d}r\wg \mathrm{d}\phi -2Q^3 J\wedge\mathrm{d}t\wedge \mathrm{d}x_1 \wedge \mathrm{d}x_2\,,\nonumber\\
&&J=-\frac{1}{6}\left(\mathrm{d}\theta_1 \wedge \sin\theta_1\mathrm{d}\phi_1 +\mathrm{d}\theta_2 \wedge \sin\theta_2\mathrm{d}\phi_2\right)\,.
\eea
To avoid singularities at the tip of the cigar, we have to require the compact $\phi$ direction to be periodic, and more precisely
\bea
\label{RQ}
\phi \sim \phi + 2\pi R\,, \quad R = \frac{2}{r^2f^\prime(r)}\bigg|_Q = \frac{1}{3Q}\,.
\eea
For a lighter notation, we have fixed the $AdS$ radius $l= (\frac{27}{4}\pi N)^{1/4} =1$. Moreover, let us evaluate the quantization condition for the self-dual five-form flux in eq.(\ref{AdS5xT11_G5}), namely
\ba
\label{QD3seed}
Q_{D3} = \frac{1}{(2\pi)^4}\int_{\widehat{T^{1,1}}}F_5 =  N,
\ea
that allows us to interpret the background as sourced by a stack of $N$ D3-branes laying at the tip of a cone in the $T^{1,1}$ manifold \cite{Klebanov:1998hh}.

\subsection{Marginal TsT deformation}\label{sec:marginal}
In this section, we are interested in deforming the $\widehat{AdS_5}\times\widehat{T^{1,1}}$ type IIB supergravity solution in eqs. (\ref{AdS5xT11}) and (\ref{AdS5xT11_G5}), in such a way to obtain a new background dual to a (marginal) beta deformation of the original field theory. In doing that we follow the strategy carried out in \cite{Lunin:2005jy} and \cite{Castellani:2024ial} by introducing the beta or TsT transformation. 
The latter, originally introduced in the context of $AdS_5\times S^5$ geometry by Lunin and Maldacena \cite{Lunin:2005jy}, is a useful technique in supergravity for generating new solutions from existing backgrounds. The process can be broken down into the following key steps:
\begin{itemize}
\item Identification of the symmetry: start by identifying a $U(1)\times U(1)$ symmetry within the isometry group of the supergravity solution. These symmetries allow shifts in two coordinates, that we denote with $\Theta_1$ and $\Theta_2$.
\item T-Duality transformation: perform a T-duality transformation along the first $U(1)$ direction, corresponding to the circle associated with $\Theta_1$.
\item Shift in the second isometry direction:
apply a shift in the second $U(1)$ direction by transforming $\Theta_2$ as follows:
\ba
\Theta_2 \longrightarrow \Theta_2 + \gamma \Theta_1\,.
\ea
Here, $\gamma$ is a real parameter that specifies the magnitude of the shift. This transformation effectively mixes the second direction, $\Theta_2$, to the first direction, $\Theta_1$.
\item T-dualization back:
finally, perform another T-duality along the first $U(1)$ direction (related to $\Theta_1$). This completes the TsT-transformation, yielding a new string theory background that is connected to the original one through these duality operations.
\end{itemize}
If the two isometries involved in the TsT are chosen within the ones of the internal manifold, \textit{i.e.} inside the R-symmetry of the theory, the resulting supergravity solution is then conjectured to be dual to a field theory deformed by the introduction of a marginal operator \cite{Lunin:2005jy,Castellani:2024ial}.\\
Let us specify to the case of $\widehat{AdS_5}\times\widehat{T^{1,1}}$. It is convenient to express the $T^{1,1}$ metric in eq. (\ref{AdS5xT11}) in the following useful form \footnote{Here, to not overweight the notation, we define $s_i = \sin\theta_i\,, c_i = \cos\theta_i$.}
\ba
\label{dsT11}
 \mathrm{d}s^2 _{T^{1,1}} &=&\frac{1}{6}\sum_{i=1}^2\mathrm{d}\theta_i^2+ \frac{s_1^2s_2^2}{324W}D\psi^2+  h \left(\mathrm{d} \phi_1 + \frac{c_1 c_2}{9h}d\phi_2 +
\frac{c_1}{ 9h} D\psi\right)^2
+ \frac{W}{h} \left( \mathrm{d}\phi_2 + \frac{ c_2 s_1^2}{54W} D\psi \right)^2\,,\nonumber\\
&&h=  \frac{c_1^2}{9}+\frac{s_1^2}{6},\qquad
W= \frac{1}{54}(c_2^2 s_1^2 + c_1^2 s_2^2 ) +\frac{s_1^2 s_2^2}{36}\,, \qquad D\psi = \mathrm{d}\psi + 3\mathcal{A}\,,
\ea
making more explicit the $\left(\phi_1\,,\phi_2\right)$ two torus  part of the metric.
Moreover, we can massage further the expression in eq. (\ref{dsT11}), writing it as
\ba
\label{dsT112}
\mathrm{d}s^2 _{T^{1,1}} &=&\frac{1}{6}\sum_{i=1}^2\mathrm{d}\theta_i^2+ \frac{s_1^2s_2^2}{324W}D\psi^2+  h \left(D\tilde\phi_1+ \frac{c_1 c_2}{9h}D\tilde \phi_2 \right)^2
+ \frac{W}{h} D\tilde\phi_2^2\,,\nonumber\\
D\tilde\phi_1 &=& \mathrm{d}\phi_1 +
\frac{c_1}{9h} \left( 1 -\frac{c_2 ^2 s_1^2}{ 54\,W}\right)D\psi\,,\nb\\
D\tilde\phi_2 &=& \mathrm{d}\phi_2 + \frac{ c_2 s_1^2}{ 54\,W}D\psi\,.
\ea

Now, following similar steps to the ones pointed out above, we can perform a TsT-transformation on the $\widehat{AdS_5}\times\widehat{T^{1,1}}$ solution, with metric as in eq. (\ref{dsT112}). 
The two-torus along which we perform the TsT transformation can be chosen as spanned by the $\phi_1$ and $\phi_2$ directions. In particular, we T-dualize and shift along $\phi_1$ and $\phi_2$, respectively.
After the beta-transformation, we end up with the following type IIB supergravity solution (in string frame)
\bea
\label{betadsT11A}
\mathrm{d}s^2_{10} &=& r^2 (-\mathrm{d}t^2+\mathrm{d}x_1^2 + \mathrm{d}x_2^2 +  f(r)\mathrm{d}\phi^2) + \frac{\mathrm{d}r^2}{ r^2 f(r)} +\nb\\
 &&+\frac{1}{6}\sum_{i=1}^2\mathrm{d}\theta_i^2+ \frac{s_1^2s_2^2}{324\,W}D\psi^2+  G\,h \left(D\tilde\phi_1+ \frac{c_1 c_2}{9h}D\tilde \phi_2 \right)^2
+ G\,\frac{W}{h} D\tilde\phi_2^2\,,\nonumber\\
B &=&\gamma W \,G D\tilde\phi_1\wg D\tilde\phi_2\,, \qquad e^{2\Phi} = G\,,\nb\\
F_5 &=& \left(1 + \star_{10\beta}\right)G_5\,,\quad F_3 =  \frac{\gamma}{G} \,\,i_{\phi_2}i_{\phi_1}\star_{10\beta}G_5\,, \quad F_7 =G_5 \wg B\,, 
\eea
where we have defined
\bea
\label{GT11}
G^{-1} = 1 + \gamma^2 W\,.
\eea
The Hodge star operator $\star_{10\beta}$ refers to the metric in eq. (\ref{betadsT11A}), while the five form $G_5$ has the form as in eq. (\ref{AdS5xT11_G5}), or more explicitly
\begin{multline}
\label{G5beta}
G_5=-4r^3 \mathrm{d}t\wg \mathrm{d}x_1 \wg\mathrm{d}x_2\wg \mathrm{d}r\wg \mathrm{d}\phi+\frac{Q^3}{3}\mathrm{d}t\wg \mathrm{d}x_1 \wg \mathrm{d}x_2\wg\left(\mathrm{d}\theta_1 \wedge \sin\theta_1\mathrm{d}\phi_1 +\mathrm{d}\theta_2 \wedge \sin\theta_2\mathrm{d}\phi_2\right)\,.
\end{multline}
We refer to the IIB supergravity solution displayed by eqs.(\ref{betadsT11A}) and (\ref{G5beta}) as the beta-transformed background. Notice that since the deformation acted in such a way that the latter background metric is modified only at the level of the internal manifold, leaving the $AdS$ one untouched, the “conformal” properties of the solution are unchanged.\footnote{Notice that here we are using an abuse of notation, since \textit{a priori} the seed soliton background in (\ref{AdS5xT11}) is not conformal.} This means that the TsT-transformed solution in eq.(\ref{betadsT11A}) then corresponds to a marginal deformation of the original dual field theory.
In analogy with eq.(\ref{QD3seed}), it is interesting to check that the D3-branes quantization condition still holds. Indeed, we have that in this case the Page charge is given by \cite{Benini:2007gx}
\ba
\label{QD3seed2}
Q^\prime_{D3} = \frac{1}{(2\pi)^4}\int_{\Sigma_5}\left[\star_{10\beta}G_5 + B\wg F_3\right] =  \frac{1}{(2\pi)^4}\int_{\Sigma_5}\star_{10\beta}G_5\left[ 1+ \gamma^2 W\right]= \frac{1}{(2\pi)^4}\int_{\Sigma_5}\star_{10}G_5 = N,\nb\\
\ea
where, the five cycle $\Sigma_5$ is spanned by the five internal directions $\left[\theta_1,\theta_2, \phi_1,\phi_2,\psi\right]$.
Furthermore, in the same spirit of \cite{Lunin:2005jy,Castellani:2024ial} we can evaluate the Page three form flux over the three cycle $\Sigma_3 = \left[\theta_1,\theta_2,\psi\right]$ as
\ba
\label{QD5}
Q^\prime_{D5} = \frac{1}{(2\pi)^2}\int_{\Sigma_3}F_3 =  \frac{1}{(2\pi)^2}\int_{\Sigma_3}\frac{\gamma}{G} \,\,i_{\phi_2}i_{\phi_1}\star_{10\beta}G_5= \frac{\gamma}{(2\pi)^4}\int_{\Sigma_5}\star_{10}G_5 =\gamma N.
\ea
This suggests that for some special values of the $\gamma$ parameter, namely those such that $\gamma N = k \in \mathbb{N}$, the results in eq.(\ref{QD5}) provide a good charge quantization condition for a novel stack of $k$ D5-branes. Finally, let us notice that there are no other additional brane sources introduced by the TsT transformation since no other compact cycles are supporting the fluxes in eq.(\ref{G5beta}).

As in the case of the beta-transformation of $AdS_5\times S^5$ and $\widehat{AdS_5}\times \widehat{S^5}$ backgrounds \cite{Castellani:2024ial} (as discussed also in \cite{Liu:2019cea}), verifying the type IIB supergravity Einstein, Maxwell and Bianchi equations for the solution in eqs. (\ref{betadsT11A}) and (\ref{G5beta}) turns out to be very requiring from the computational point of view. Nevertheless, the solution is expected to work properly.

\subsection{Dipole TsT transformation}\label{sec:dipole}
In this subsection, we perform an alternative TsT transformation on the $\widehat{AdS_5}\times \widehat{T^{1,1}}$ background of eqs.(\ref{betadsT11A}) and (\ref{G5beta}) by setting the two torus as having a $U(1)$ direction along the internal manifold $T^{1,1}$ and the other one along the deformed $AdS_5$ spacetime. In doing so we will choose the two directions as the ones describing the compact cycle $\phi$ and the angular $\psi$ in eq.(\ref{AdS5xT11}).\footnote{Remarkably, notice that since the TsT transformation involves the $\phi$ direction, on which the SUSY spinor depends, the T-dualization process here breaks all the supersymmetry.}\\
It is useful to rewrite the ten-dimensional metric in eq.(\ref{AdS5xT11}) as 
\ba
\label{T11dip}
& &\mathrm{d}s^2_{10}=\mathrm{d}s^2_{8} + r^2f(r)\mathrm{d}\phi^2+  \frac{1}{9}\left(\mathcal{D}\psi +3\mathcal{A}\right)^2\,,\nb\\
&& \mathrm{d}s^2_{8} = r^2 (-\mathrm{d}t^2+\mathrm{d}x_1^2 + \mathrm{d}x_2^2 ) + \frac{\mathrm{d}r^2}{ r^2 f(r)} +\frac{1}{6}\sum_{i=1}^2\left(\mathrm{d}\theta_i^2+\sin^2\theta_i\mathrm{d}\phi_i^2\right)\,,
\ea
where we also defined
\ba
&& \mathcal{D}\psi  = \mathrm{d}\psi + \sum_{i=1}^2\cos\theta_i\mathrm{d}\phi_i\,.
\ea
Looking at the prescriptions for the TsT transformation in Subsection \ref{sec:marginal}, we can T-dualize with respect to the R-symmetry direction $\psi$, then apply the shift $\phi\to \phi +\gamma \psi$ and finally T-dualize back along $\psi$. Thus, we get that the TsT-deformed metric is given by
\ba
\label{T11dip}
& &\mathrm{d}s^2_{10}=\mathrm{d}s^2_{8} + 9 G r^2f(r)\mathrm{d}\phi^2+  G\left(\mathcal{D}\psi +3\mathcal{A}\right)^2\,,\nb\\
&& G^{-1} = 9 + \g^2r^2f(r)\,,
\ea
along with the field content
\ba
\label{FieldDipoleT11}
&&B =\gamma r^2 f(r) \,G\,\mathcal{D}\psi \wg \mathrm{d}\phi\,, \qquad  \quad e^{2\Phi} = 9 \,G\,,\nb\\
&&F_5  = \left(1 + \star_{10\beta}\right)G_5\,,\quad F_3 =  -\frac{\gamma Q^3}{9r^3} \,\,\mathrm{d}r\wg \left[\sin\theta_1 \mathrm{d}\theta_1 \wg\mathrm{d}\phi_1+\sin\theta_2 \mathrm{d}\theta_2 \wg\mathrm{d}\phi_2\right]\,, \quad F_7 =G_5 \wg B\,.\nb\\
\ea
As in \cite{Bergman:2001rw} and \cite{Castellani:2024ial}, we refer to the solution in eqs.(\ref{T11dip}) and (\ref{FieldDipoleT11}) as the dipole-deformed $\widehat{AdS_5}\times \widehat{T^{1,1}}$. The TsT transformation applied along the $\phi$ direction, non-trivially alters the $AdS$ structure of the solution. Specifically, the metric in eq.(\ref{T11dip}) no longer exhibits an $AdS_5$ asymptote in the UV (large $r$ limit). This dipole-deformation corresponds to the introduction of an irrelevant operator in the dual field theory, which modifies its ultraviolet behavior \cite{Castellani:2024ial}.\\
The quantization of Page fluxes closely follows the observations made in Subsection \ref{sec:marginal}. Using the expressions for the fluxes in eq.(\ref{FieldDipoleT11}), we obtain:
\ba
Q_{D3}^\prime = \frac{1}{(2\pi)^4}\int_{\Sigma_5}\star_{10\beta} G_5 = \frac{1}{(2\pi)^4}\int \star_{10} G_5 = N \,,
\ea
with the five-cycle defined as $\Sigma_5 = \left[\theta_1,\theta_2,\phi_1,\phi_2, \psi\right]$. Notice that in this case there is no contribution from the $B$ and $F_3$ fields: in fact, $F_3$ extends only along non-compact directions. For the same reason, we cannot impose a charge quantization condition for D5-branes: given that $F_3$ has no magnetic components on a compact cycle, we conclude that in the present dipole-deformed background (including the case with rational $\gamma$), there are no D5-branes present.

We can investigate the smoothness of this background in both the IR and UV limits, specifically for $r$ near the tip of the cigar and $r\to \infty$, respectively. In doing that it is useful to examine the Ricci scalar of the metric in eq.(\ref{T11dip}), which is given by
\ba
\label{RicciT11dipole}
R = \frac{2 \gamma ^2 \left(Q^{12} \left(26 \gamma ^2 r^2+252\right)+Q^6 \left(18 r^6-34 \gamma ^2 r^8\right)+r^{12} \left(8 \gamma ^2 r^2+135\right)\right)}{\left(-\gamma ^2 Q^6 r+\gamma ^2 r^7+9 r^5\right)^2}\,.
\ea
Notice that the expression in eq.(\ref{RicciT11dipole}) in the IR and UV limits approaches constant values, \textit{i.e.}
\ba
\label{T11riccitip}
R\vert_{r= Q} = 10 \gamma ^2 Q^2\,,   \quad R\vert_{r\to \infty} = 16-\frac{18}{\gamma ^2 r^2}+O\left(r^{-4}\right)\,,
\ea
suggesting the non-singular behavior of the geometry.\\
In addition to eq.(\ref{T11riccitip}) we can also compute in the same limits the contraction of the Ricci tensor and the  Kretschmann scalar. In particular, these are given by
\ba
&& R_{\m\n}R^{\m\n}\vert_{r=Q} = 4 \left(9 \gamma^4 Q^4-2 \gamma^2 Q^2+54\right)\,,\quad R_{\m\n}R^{\m\n}\vert_{r\to \infty}=  176-\frac{864}{\gamma ^2 r^2}+O\left(r^{-4}\right)\,,\nb\\
&& R_{\m\n\r\s}R^{\m\n\r\s}\vert_{r=Q} = 4(11 \gamma^4 Q^4+40 \gamma^2 Q^2+150)\,,\quad R_{\m\n\r\s}R^{\m\n\r\s}\vert_{r\to \infty}= 328-\frac{2232}{\gamma ^2 r^2}+O\left(r^{-4}\right)\,,\nb\\
\ea
which are again finite. \\ 
For the present dipole-deformed solution in eqs.(\ref{T11dip}) and (\ref{FieldDipoleT11}), we used Mathematica to verify the type IIB supergravity Einstein, Maxwell, and dilaton equations, along with the Bianchi identities for the NS and RR fields.

\section{Field theory and observables}\label{QFT-section}
In this section, we analyze holographic observables for the QFTs corresponding to the TsT deformed backgrounds introduced in Subsections \ref{sec:marginal} and \ref{sec:dipole}. We compare them with the respective ones evaluated in the seed $\widehat{AdS_5}\times \widehat{T^{1,1}}$ background to infer their dependence on the massive KK modes physics.\\ 
The type IIB supergravity seed-solution $\widehat{AdS_5}\times \widehat{T^{1,1}}$ presented in \cite{Chatzis:2024top,Chatzis:2024kdu} (see eqs.(\ref{AdS5xT11})-(\ref{AdS5xT11_G5})) is holographically dual to a twisted (SUSY preserving) compactification of the Klebanov-Witten SCFT \cite{Klebanov:1998hh} (for a review, see \cite{Herzog:2002ih}). The latter corresponds to the low-energy limit of the worldvolume theory on a stack of $N$ D3 branes posed at the apex of a Ricci-flat 6d cone whose base is the 5d Einstein manifold $T^{1,1}$. In particular, this theory is a $\mathcal{N}=1$ SUSY $SU(N) \times SU(N)$ gauge theory coupled to four chiral superfields, $A_i$ ($i=1,2$) and $B_j$ ($j=1,2$), which transform in the $(N, \overline{N})$  and $(\overline{N}, N)$ representations, respectively,\footnote{The $A_i$ fields form a doublet under a “left” global $SU(2)$ symmetry, while the $B_i$ fields form a doublet under a “right” global $SU(2)$. In addition, there is also another $U(1)$ R-symmetry under which all the chiral superfields have charge $1/2$. These symmetries fix completely the form of the superpotential in eq.(\ref{superpotW}) \cite{Klebanov:1998hh}.} flowing to a non-trivial infrared fixed point perturbed with a marginal superpotential of the form \footnote{As stated in \cite{Lunin:2005jy} a marginal TsT deformation of the type IIB supergravity $AdS_5\times T^{1,1}$ solution is conjecture to be dual to a marginal deformed Klebanov-Witten SCFT given by a modification of superpotential in eq.(\ref{superpotW}) of the form: $\Tr\left[A_1B_1A_2B_2-A_2B_1A_1B_2\right]\rightarrow \Tr\left[e^{i\pi\g}A_1B_1A_2B_2-e^{-i\pi\g}A_2B_1A_1B_2\right]$ \cite{Benvenuti:2005wi}. }
\ba
\label{superpotW}
W \sim \e^{ij}\e^{kl}\Tr\left[A_iB_kA_jB_l\right]\,.
\ea
Remarkably, this CFT does not admit a weakly coupled regime. This feature makes it very challenging to compute observables in the original field theory and also in its compactified and deformed versions. As a result, the holographic approach provides a helpful tool for making predictions about them.

\subsection{Observables}
\label{sec:observables}
In this subsection, we explore various interesting observables in the context of the quantum field theories dual to the TsT-deformed Type IIB backgrounds introduced in Section \ref{section-geometry}. We provide a brief overview of the main characteristics of the discussed quantities. For a more comprehensive and detailed analysis of similar observables in the field theories dual to TsT-transformed $\widehat{AdS_5} \times \widehat{S^5}$ solutions, we refer to \cite{Castellani:2024ial}.
\subsubsection{Wilson loops}
One of the key features of the TsT-transformed $\widehat{AdS_5}\times \widehat{T^{1,1}}$ backgrounds in Section \ref{section-geometry} that we want to study is their confinement behavior. To explore the confining properties, we compute the potential energy for a (quenched) quark-anti-quark pair via the holographic rectangular Wilson loop, identified with the boundary of a probe open string in the bulk geometry \cite{Rey:1998ik,Maldacena:1998im}.
Let us consider a classical string of length $L$. Its embedding is set as
\ba
\label{embedding}
t = \t \,, \quad x^1 = \s\,, \quad r = r(\s)\,, \quad \s \in \left[-L/2, L/2\right]\,,
\ea
with the other coordinates held constant. Since the marginal and dipole TsT transformations do not affect the (3d) Minkowski or radial coordinates, the confining behavior remains the same as the original background in eq.(\ref{AdS5xT11}), as the relevant Wilson loop is unaffected by the deformation.

For both backgrounds, the induced worldsheet metric from the embedding in eq.(\ref{embedding}) is
\ba
\mathrm{d}s^2_{\text{ind}} = r^2\left[-\mathrm{d}\t^2 + \left(1 + \frac{r'(\s)^2}{r^4 f(r)}\right)\mathrm{d}\s^2\right], \quad r^\prime(\s) = \partial_\s r(\s)\,,
\ea
leading to the Nambu-Goto action for the probe string
\ba
\label{SNG_Wilson}
S_{NG} = \frac{T}{2\pi} \int_{-L/2}^{L/2} \mathrm{d}\sigma \sqrt{F_w^2(r) + G_w(r)^2 r'(\sigma)^2}\,,\quad F_w(r) = r^2 \,,\quad G_w(r) = \frac{1}{\sqrt{f(r)}}\,.
\ea
From eq.(\ref{SNG_Wilson}), we derive the effective potential \cite{Nunez:2009da,Chatzis:2024top} 
\ba
V_{\text{eff}}(r)=  \frac{r^2(\s)}{r_0^2} \sqrt{f(r)\left(r^4(\s) - r_0^4\right)},
\ea
where $r_0$ is the turning point of the string in the radial direction. 
Notice that the divergent and vanishing behaviors of $V_{\text{eff}}(r)$ as $r\to \infty$ and $r\to r_0$, respectively, indicate (as expected) confinement \cite{Nunez:2009da}. The quark-anti-quark separation length is given by
\ba
L_{QQ}(r_0) = 2r_0^2 \int_{r_0}^{\infty} \frac{r \, \mathrm{d}r}{\sqrt{(r^6 - Q^6)(r^4 - r_0^4)}}\,,
\ea
which diverges as $r_0 \to Q$, signaling confinement as the string deeper probes the IR region, leading to infinite quark separation. Finally, the quark-anti-quark (regularized) energy is
\ba
\label{EQQ}
E_{QQ}(r_0) = r_0^2 L_{QQ}(r_0)+ 2 \int_{r_0}^{\infty}\mathrm{d}r \,\frac{\sqrt{r^4-r_0^4}}{r^2 \sqrt{f(r)}}- 2 \int_{Q}^{\infty}\mathrm{d}r \,\frac{1}{ \sqrt{f(r)}}\,,
\ea
with the linear dependence of $E_{QQ}$ on  $L_{QQ}$ in eq.(\ref{EQQ}) further indicating confinement \cite{Rey:1998ik,Maldacena:1998im}. Additionally, it is possible to show that the string embedding is expected to be stable \cite{Castellani:2024ial,Chatzis:2024top,Chatzis:2024kdu}.

In conclusion, both the marginal and dipole TsT-transformed backgrounds in Subsections \ref{sec:marginal} and \ref{sec:dipole} exhibit confining behavior, consistently with the seed $\widehat{AdS_5}\times \widehat{T^{1,1}}$ background.

\subsubsection{'t Hooft loops}
Another relevant and interesting observable to analyze in the background of Subsections \ref{sec:marginal} and \ref{sec:dipole} is the 't Hooft loop. The system of a magnetic monopole-anti-monopole pair is accounted holographically by a probe D3-brane extended in the $\xi  \in [t,x_1, \phi, \alpha]$ directions (where $\alpha$ is a direction along $\widehat{T^{1,1}}$), with the radial direction as a function of $x_1$, \textit{i.e.} $r(x_1)$. The remaining coordinates are fixed. The DBI action for the D3-brane probe is
\ba
\label{DBI}
S_{D_3}= T_{D3}\int{}\mathrm{d}^{4} \xi\,e^{-\Phi} \sqrt{-\det\left(P[g]_{ab} + P[B]_{ab} \right)}\,,
\ea
where $P[\cdot]$ represents the pull-back of the fields on the worldvolume. \\ 
After integrating the $[\phi, \alpha]$ directions, we obtain a two-dimensional effective action similar to the one for the Wilson loop in eq.(\ref{SNG_Wilson}).\\
We now consider the TsT-transformed background of Subsection \ref{sec:marginal}, and take the D3 probing brane as laying on the directions $[t,x_1, \phi, \psi]$. Setting for convenience $\theta_1 =0$ and $\theta_2 = \pi/2$, the induced metric and dilaton are given by
\ba
\label{D3ind}
&&\mathrm{d}s^2_{D3} = r^2\left[-\mathrm{d}t^2 + \left(1+ \frac{r'(x_1)^2}{r^4f(r)}\right)\mathrm{d}x_1^2 + f(r)\mathrm{d}\phi^2 \right] + \frac{G}{9}\left(\mathrm{d}\psi + 3\mathcal{A}\right)^2\,, \nb\\
&& e^{2\Phi} =G\,, \quad G^{-1}= 1+ \frac{\gamma^2}{54}\,.
\ea
In this case, the $B$ field does not contribute. The DBI action reduces to
\ba
\label{SeffHL}
S_{\text{eff}} = T_{D3}L_{\phi} L_{\psi} \frac{T}{3} \int_{-L/2}^{L/2} \mathrm{d}x_1\, \sqrt{F_t(r)^2 + G_t(r)^2 r'^{\,2}}\,,\quad F_t(r) = r^3 \sqrt{f(r)}\,,\quad G_t(r) = r\,.
\ea
Since the function $F_t(r)$ vanishes near the tip, we observe that the effective string tension goes to zero as $r \to Q$, indicating that in the IR limit the magnetic monopoles can be separated without any energy cost, suggesting a screening behavior.\\
The effective potential and monopole-anti-monopole separation length $L_{MM}(r_0)$ are
\ba
\label{VeffHL}
V_{\text{eff}}(r) = \frac{F_t(r)}{F_t(r_0) G_t(r)} \sqrt{F_t(r)^2 - F_t(r_0)^2}\,,
\ea
and
\ba
\label{LMM}
L_{MM}(r_0) = 2F_t(r_0) \int_{r_0}^{\infty} \frac{\mathrm{d}r}{r^2\sqrt{f(r)}\sqrt{r^6 f(r) - F_t(r_0)^2}}\,.
\ea
As $r_0 \to Q$, $L_{MM}$ approaches zero, further confirming the screening. The energy is
\ba
E_{MM}(r_0) = F_t(r_0) L_{MM}(r_0) + 2 \int_{r_0}^{\infty} \frac{\sqrt{r^6 f(r) - F_t(r_0)^2}}{r^2 \sqrt{f(r)}}-2\int_{Q}^{\infty}\mathrm{d}r\, r\,,
\ea
and is regularized by adding a counterterm, as in the Wilson loop case. Notice that the linear term vanishes in the near-tip limit.

Next, we consider the dipole-deformed background in Subsection \ref{sec:dipole}. The probe D3-brane this time is embedded along the $[t,x_1,\phi,\varphi]$ coordinates, with again $r(x_1)$. After integrating the compact coordinates, the effective action turns out to be equal to the one in eq.(\ref{SeffHL}), showing the same screening behavior for monopole-anti-monopole systems. Interestingly, similarly to what was observed in \cite{Castellani:2024ial}, the dilaton and $B$ field contributions are fundamental in canceling out the $G$ terms, leading to the same results as in the original background \cite{Chatzis:2024top,Chatzis:2024kdu}. Finally, we refer to \cite{Chatzis:2024top,Chatzis:2024kdu,Castellani:2024ial} for a detailed analysis of the 't Hooft loops and their stability in the seed and TsT-deformed $\widehat{AdS_5}\times \widehat{S^5}$ backgrounds.

\subsubsection{Entanglement Entropy}\label{sec:EE}

In this subsection, we want to study holographically the Entanglement Entropy (EE) of a strip of length $L$. Following the holographic procedure (see \textit{e.g.} \cite{Ryu:2006bv,Klebanov:2007ws,Kol:2014nqa}), the EE for two boundary regions can be evaluated by considering the minimal-area codimension-two manifold $\Sigma_8$ that attaches to the boundary along the entangling surface. As noted in \cite{Klebanov:2007ws,Kol:2014nqa}, the entangling surface can be taken as a strip of length $L$. The action to be minimized is expressed in terms of the determinant of the induced metric on $\Sigma_8$ and exponentials of the dilaton as
\ba\label{S_EE}
    S_{\mathrm{EE}} = \frac{1}{4G_N} \int_{\Sigma_8} \mathrm{d}^8x \, e^{-2\Phi} \sqrt{\mathrm{det} g_{\Sigma_8} }
\ea
where $G_N$ is the ten-dimensional Newton's constant.\\
Let us consider the surface $\Sigma_8$ chosen as a fixed time surface that extends along the internal $\widehat{T^{1,1}}$ submanifold and the $[x_1,x_2,\phi]$ AdS directions, with $r = r(x_1)$. As noticed in \cite{Castellani:2024ial}, using the deformed metrics and dilatons in eqs.(\ref{betadsT11A}), (\ref{T11dip}) and (\ref{FieldDipoleT11}), we can observe that the relevant quantities involved in eq.(\ref{S_EE}) transform under TsT as
\ba
\mathrm{det} g_{\Sigma_8} \quad \longrightarrow \quad G^2 \mathrm{det} g_{\Sigma_8}\,,\quad e^{2\Phi_0} \quad \longrightarrow \quad e^{2\Phi_0} G\,.
\ea
Thus, we conclude that in both the marginal and dipole cases, the action in eq.(\ref{S_EE}) reduces to the one of the seed solution in eq.(\ref{AdS5xT11}). The latter exhibits a first-order phase transition between the connected and disconnected configurations \cite{Chatzis:2024kdu,Chatzis:2024top} which is interpreted as a signal for confinement.\footnote{Remarkably, despite the dipole deformed solution of Subsection in \ref{sec:dipole} no longer has a local field description in the UV, the EE still returns a field theory result. This feature suggest us to see the dipole TsT transformation as a softer form of a non-commutative deformation.} 
Finally, it is interesting to emphasize on the fact that the TsT deformed backgrounds presented in Section \ref{section-geometry}, display the same key features regarding relevant field theory observables as Wilson loops, 't Hooft loops and Entanglement Entropy as those found \cite{Chatzis:2024top,Chatzis:2024kdu,Castellani:2024ial}. This property points to a sort of universality underlying the corresponding dual field theories.
\subsubsection{Holographic Central Charge Flow}\label{sec:Hol_central_charge}
Another observable of interest that we want to analyze is the holographic central charge. This is a significant quantity in supergravity solutions dual to conformal field theories (even though this is not the case of the QFTs dual to backgrounds presented in Section \ref{section-geometry}) and reflects the number of degrees of freedom of the latter. In holography, we can study the central charge flow, which is a monotonic function representing the counting of degrees of freedom across dimensions \cite{Macpherson:2014eza,Bea:2015fja}. To define this flow, we express the metric and the dilaton of a general background dual to a $(d+1)$-dimensional QFT as follows
\ba
ds^{2} &=&-\a_0\mathrm{d}t^2+ \sum_{n =1}^{d} \a_n \mathrm{d}x_n^2 +\left(\prod_{n =1}^d \alpha_n \right)^{\frac{1}{d}}\b \left( r\right) dr^{2} +g_{ij} (dy^{i}-A^i)(dy^{j}-A^j)\,,\nb\\
\Phi &=& \Phi(r, y^i)\,
\label{conventions central charge}
\ea
In this expression, $y^{i}$ are the coordinates of the internal manifold, while the metric on the submanifold extending in the $\left[x_n,y^i\right]$ directions is given by
\ba
\hat g_{ab}\mathrm{d}\xi^a \mathrm{d}\xi^b \equiv \sum_{n =1}^{d} \alpha_n \mathrm{d}x_n^2 +g_{ij} dy^{i}dy^{j}\,.
\ea
Then, we introduce the function 
\ba
\label{Hcflow}
H^{\frac{1}{2}}(r) \equiv \int\mathrm{d}\xi^a \, e^{-2\Phi }\sqrt{\det \hat g_{ab}}\,.
\ea
Following \cite{Macpherson:2014eza,Bea:2015fja}, the holographic central charge flow is given by
\ba
c_{\text{flow}}=d^{d}\frac{\beta \left( r\right) ^{d/2}H^{\left( 2d+1\right) /2}}{G^{(10)}_{N}\left( H^{\prime }\right) ^{d}}\,.
\label{chol}
\ea
Let us specialize to the case of backgrounds in Section \ref{section-geometry} and set $d=3$. The specific forms of the functions in eq. (\ref{conventions central charge}) and the dilaton depend on the chosen background. 
In the case of the marginal deformed solution in eq.(\ref{betadsT11A}), using the conventions from eq. (\ref{conventions central charge}), we find
\ba
\label{cflow_alpha}
\a_{1,2} = r^2\,, \quad \a_{3} = r^2 f(r)\,,\quad \beta(r) =\frac{1}{r^4}f(r)^{-\frac{4}{3}}\,.
\ea
Moreover, we can evaluate $H(r)$ from eq. (\ref{Hcflow}), yielding \footnote{Notice that similarly to the EE in eq.(\ref{S_EE}), the function $H$ in eq.(\ref{H2cflow}) is invariant under the TsT-deformation.}
\ba
\label{H2cflow}
H^{\frac{1}{2}}(r)= \text{vol}_{T^{1,1}}r^3 \sqrt{f(r)}\,.
\ea
Then, we find
\ba
\label{c2hol}
c_{\text{flow}} = \frac{\text{vol}_{T^{1,1}}}{8G^{(10)}_{N}}\left(1-\frac{Q^6}{r^6}\right)^{\frac{3}{2}}\,.
\ea
Notably, in the IR limit ($r\to Q$), $c_{\text{flow}}$ vanishes, indicating a lack of dynamical degrees of freedom (a gapped system). On the other hand, in the UV regime, $c_{\text{flow}}$ approaches a non-zero value given by
\ba
\label{cUV}
c_{\text{UV}} = \frac{\text{vol}_{T^{1,1}}}{8G^{(10)}_{N}}\,.
\ea
Thus, we observe that $c_{\text{flow}}$ flows monotonically between an IR gapped 3d system and a UV 4d SCFT point, where it exhibits a value proportional to the volume of the internal manifold as in eq. (\ref{cUV}). Remarkably the latter behavior coincides with the one of the seed background $\widehat{AdS_5}\times\widehat{T^{1,1}}$ \cite{Chatzis:2024kdu}.\footnote{Actually, this feature is similar to all the different solutions studied in \cite{Chatzis:2024top,Chatzis:2024kdu}, suggesting a form of universal character of this observable.} This result was somehow expected since we are marginally deforming the dual field theory.
On the other hand, this is not the case of the dipole deformation in the background of eq.(\ref{T11dip}). While $\alpha_1,\alpha_2$ remain the same as in eq. (\ref{cflow_alpha}), we find
\ba
\a_1=\a_2=r^2\,,\quad \a_3=r^2 f(r)G(r)\,\quad \b(r)= \frac{1}{r^4 f(r)^{4/3} G^{1/3}}\,.
\ea
These imply that upon applying eq. (\ref{chol}), we obtain
\ba
 c_{flow}=   \frac{\text{vol}_{T^{1,1}}}{8G^{(10)}_{N}}\left(1-\frac{Q^6}{r^6}\right)^{\frac{3}{2}} \times \frac{1}{\sqrt{G(r)}}\,.
\ea
While this expression still indicates a gapped phase in the IR, the UV behavior is characterized by an unbounded $c_{flow}$. This aligns with our expectations, given that the $r\to\infty$ asymptotics of the background is not $AdS_5$ and the deformation corresponds to the introduction of an irrelevant operator in the dual field theory.
This implies the necessity for this QFT to have a UV completion.\footnote{A UV description could be possibly found in terms of string theory, in the same lines as in \cite{Nunez:2023nnl},\cite{Nunez:2023xgl}.}

\subsubsection{\texorpdfstring{$\mathbb{Z}_N$}{ZN} symmetry broken and unbroken phases}
In \cite{Kumar:2024pcz,Castellani:2024ial} the analysis of the broken and unbroken phases of a $\mathbb{Z}_N$ symmetry in 4d twisted compactified field theories is presented. Following a similar approach, here we want to examine the symmetry breaking process of the latter symmetry using as order parameter a Wilson loop of the $SU(N)$ gauge field along the compactified $S^1$ cycle, \footnote{In the case in which the Wilson loop wraps along the (Euclidean) temporal direction, it turns out to be an order parameter for both the symmetry breaking the “electric” $\mathbb{Z}_N$ center of the $SU(N)$ gauge group and confinement \cite{tHooft:1977nqb,Witten:1998zw}.} namely
\ba
\label{Polyakov_loop}
    P_\phi =\frac1N{\rm Tr} \,\exp i\oint_{S^1_\phi}\, A^{(0)}_\phi\,,
\ea
where $A^{(0)}_\phi$ is the Kaluza-Klein zero-mode of the $\phi$ component of the gauge field after the reduction over the circle. Remarkably, as it is stated in \cite{Kumar:2024pcz}, a vanishing expectation value for $P_\phi$ in eq. (\ref{Polyakov_loop}) corresponds to an unbroken $\mathbb{Z}_N$ symmetry, while a non-vanishing one signals the spontaneous breaking of this symmetry.
The holographic computation of the latter goes through the evaluation of the on-shell Nambu-Goto action for the open string whose worldsheet approaches the Wilson loop contour at the boundary \cite{Witten:1998zw} as $\langle P_\phi\rangle \sim \,\exp\left(-S^{\text{reg}}_{NG}\right)$.\\
Starting with the marginal TsT-transformed $\widehat{AdS_5}\times \widehat{T^{1,1}}$ background of Subsection \ref{sec:marginal}, we use the following probe string Euclidean embedding
\ba 
\label{PLstring_marginal} 
\phi(\tau) = \tau,\quad r(\sigma) = \sigma\,,
\ea 
and the other coordinates are taken as constant. From eqs.(\ref{betadsT11A}) and (\ref{PLstring_marginal}) we can derive the induced metric
\ba 
\mathrm{d}s^2_{\text{ind}} =\frac{1}{r^2f(r)} \mathrm{d}r^2 + \left[r^2f(r) + G\left(1+\frac{\gamma^2}{36}\sin^2\theta_1\sin^2\theta_2\right)Q^6 \left(\frac{1}{r^2}-\frac{1}{Q^2}\right)^2\right]\mathrm{d}\phi^2. \ea 
After having introduced the dimensionless coordinate $r = Q \rho$ 
, the UV regularized on-shell Nambu-Goto action is computed as
\ba
\label{SF1_beta}
S^{\text{reg}}_{NG}
&=&\lim_{\L\to \infty}
RQ\left[\int_{1}^\Lambda\mathrm{d}\r \sqrt{1+ G\, \left(1+\frac{\gamma^2}{36}\sin^2\theta_1\sin^2\theta_2\right)\left(1+ \frac{(\r^2-1)^2}{(\r^6-1)}\right)}-\Lambda\right]\,.
\ea
The dipole-deformed case, introduced in Subsection \ref{sec:dipole}, follows a similar structure. Here, the induced metric is
\ba
\mathrm{d}s^2_{\text{ind}} =\frac{1}{r^2f(r)} \mathrm{d}r^2 + 9G(r)\left[r^2f(r) + Q^6 \left(\frac{1}{r^2}-\frac{1}{Q^2}\right)^2\right]\mathrm{d}\phi^2\,,
\ea
and the regularized on-shell action becomes 
\begin{multline}
\label{SF1reg}
S^{\text{reg}}_{NG} 
=\lim_{\L\to \infty}\,
RQ\left[\int_{1}^\L\mathrm{d}\r \sqrt{\frac{\r^4}{\r^4 + \g^2 Q^2\ell^4 (\r^6-1)}\left(1+ \frac{(\r^2-1)^2}{(\r^6-1)}\right)}-\frac{1}{\g Q\ell^2}\log \L\right]\,.
\end{multline}
Interestingly, both the marginal and dipole-deformed theories exhibit non-vanishing Polyakov-like loop expectation values, implying spontaneous $\mathbb{Z}_N$ symmetry breaking, that depend explicitly on the deformation parameter $\gamma$, reflecting the contribution of KK modes in the dynamics.

\section{Conclusions and future lines of study}\label{Conclusions}
In this paper, we investigated novel TsT-transformed (T-duality, coordinate shift, T-duality) type IIB supergravity geometries from a seed SUSY soliton $\widehat{AdS_5}\times\widehat{T^{1,1}}$ presented in \cite{Chatzis:2024kdu,Chatzis:2024top}. We started by reviewing the original geometry and the TsT-generating solutions procedure. Then we applied this transformation, constructing two novel type IIB supergravity backgrounds with distinct features. The beta TsT-deformation (involving a $U(1)\times U(1)$ toroidal symmetry inside the $T^{1,1}$ isometries) leads to a marginal deformed dual field theory, while the dipole deformation (with $U(1)\times U(1)$ symmetry associated with an internal direction and the compact circle isometries) introduces more drastic changes that modify the asymptotic structure of the spacetime and the ultraviolet (UV) behavior of the dual field theory.

Through holographic techniques, we analyzed a variety of observables in the quantum field theories dual to these deformed backgrounds, including Wilson loops, 't Hooft loops, Entanglement Entropy (EE), and the holographic central charge flow. These quantities provide crucial insights into the key properties of the dual system such as confinement, magnetic monopole screening behavior, and the energy-dependence of the number of degrees of freedom. In particular, in the calculation of the Wilson loop and the EE, we observed that the effects of the marginal and dipole deformations notably cancel, providing the same results as for the seed background: indeed, the linear scaling of the heavy quark-anti-quark pair energy with their separation length and the first order transition exhibited by the EE suggest a confining behavior for the system. Moreover, the fact that these observables coincide in the original and deformed solutions leads us to conclude that they do not receive contributions from the massive Kaluza-Klein (KK) sector.
The cancellation of the deformation effects occurs in the analysis of the 't Hooft loop as well, and the findings support the argument for screening of magnetic monopoles.
\\
Additionally, we explore the spontaneous breaking of a $\mathbb{Z}_N$ symmetry in the dual field theories using a circular (Polyakov-like) Wilson loop as an order parameter. Both in the marginal and dipole cases, the (non-vanishing) order parameter value depends non-trivially on the $\gamma$ parameter ruling the TsT-transformation. Hence, this observable is influenced by the KK modes dynamics.

Finally, we studied the measure of degrees of freedom in the systems and its flow from the UV to IR regimes, through the holographic central charge flow. On one hand, in the marginal case (as expected) this quantity is invariant under the deformation and ranges from a zero value in the IR (gapped phase) to the typical one for a CFT (proportional to the internal space volume) in the UV. On the other hand, in the dipole case, while we still observe a vanishing amount of degrees of freedom in the IR, in the UV the holographic central charge diverges, underlying the fact that the (irrelevant) dipole deformation afflicts the ultraviolet behavior of the theory. Moreover, the latter result is dependent on the UV structure of the system and the KK massive modes.

This work opens the way to several possible developments and future projects, such as:

\begin{itemize}
    \item TsT Transformations and Yang-Baxter Deformations: a promising research direction would be to apply TsT transformations and more general Yang-Baxter deformations as in \cite{vanTongeren:2015uha,vanTongeren:2016eeb,Araujo:2017jkb}, to systems similar to the ones analyzed here. This could further explore the interaction between massless and KK modes in dynamical observables of reduced (SUSY) field theories.
\item One-loop Corrections: another valuable direction would be to investigate one-loop quantum corrections to the Wilson loop configurations in the present backgrounds, and explore their (expected) dependence on UV effects. 
\item Hagedorn Behavior: the study of the (potential) Hagedorn limit of the QFTs considered here, based on recent works on the Hagedorn Temperature in confining holographic models \cite{Bigazzi:2023oqm, Bigazzi:2023hxt, Bigazzi:2024biz, Bigazzi:2024sjy}, could offer insights into their thermal properties and confinement/deconfinement phase transitions.
\end{itemize}

This work aimed to investigate novel TsT-transformed supergravity geometries and key features of their field theory duals, to contribute to a better understanding of how quantum field theories respond to marginal and dipole (irrelevant) deformations, especially in energy regimes where direct calculations are difficult or even prohibitive.

\section*{Acknowledgments} 
We are grateful to Carlos Nunez for enlightening discussions and collaboration at the beginning of this project.
For discussions, comments on the manuscript, and for sharing their ideas with us, we also wish to thank: 
Francesco Bigazzi, Aldo Cotrone and Ricardo Stuardo.
F.C. wants to thank the Swansea University for its hospitality during the research period that led to this work.


\appendix

\bibliographystyle{JHEP}
\bibliography{Ref.bib}

\end{document}